\documentclass[journal,comsoc]{IEEEtran}
\usepackage[letterpaper, left=0.653 in, right=0.653 in, top=0.71 in, bottom=1.05 in]{geometry}
\usepackage{cite} 
\usepackage{booktabs} 
\usepackage{graphicx}
\usepackage{caption}
\usepackage{subcaption}
\usepackage{amsmath,amsfonts}
\usepackage{array}
\usepackage{textcomp}
\usepackage{helvet}
\usepackage{stfloats}
\usepackage{verbatim}
\usepackage{subscript} 
\usepackage[caption=false,font=normalsize,labelfont=sf,textfont=sf]{subfig}
\usepackage{multirow}
\usepackage{algorithmic}
\usepackage{algorithm} 
\begin{document}
\setlength{\columnsep}{0.245 in}

\title{BiCSI: A Binary Encoding and Fingerprint-Based Matching Algorithm for Wi-Fi Indoor Positioning}

\author{Pei Tang,~\IEEEmembership{Student Member, IEEE}, 
        Jingtao Guo,~\IEEEmembership{Student Member, IEEE}, 
        \\and Ivan Wang-Hei Ho,~\IEEEmembership{Senior Member, IEEE}
\thanks{}}

\maketitle
\thispagestyle{empty} 

\begin{abstract}
Traditional global positioning systems often underperform indoors, whereas Wi-Fi has become an effective medium for various radio sensing services. Specifically, utilizing channel state information (CSI) from Wi-Fi networks provides a non-contact method for precise indoor positioning; yet, accurately interpreting the complex CSI matrix to develop a reliable strategy for physical similarity measurement remains challenging. This paper presents BiCSI, which merges binary encoding with fingerprint-based techniques to improve position matching for detecting semi-stationary targets. Inspired by gene sequencing processes, BiCSI initially converts CSI matrices into binary sequences and employs Hamming distances to evaluate signal similarity. The results show that BiCSI achieves an average accuracy above 98\% and a mean absolute error (MAE) of less than three centimeters, outperforming algorithms directly dependent on physical measurements by at least two-fold. Moreover, the proposed method for extracting feature vectors from CSI matrices as fingerprints significantly reduces data storage requirements to the kilobyte range, far below the megabytes typically required by conventional machine learning models. Additionally, the results demonstrate that the proposed algorithm adapts well to multiple physical similarity metrics, and remains robust over different time periods, enhancing its utility and versatility in various scenarios.
\end{abstract}


\section{Introduction}

\IEEEPARstart{D}{ue} to the high adoption in indoor access networks and greater flexibility within the unlicensed spectrum, Wi-Fi has been widely used in radio sensing services, including healthcare \cite{yang_2022_artificial, salem_2021_markov}, human activity recognition \cite{yang_2018_carefi,zhang_2022_wifi}, crowd counting \cite{belalkorany_2021_counting,jiang_2024_pacount}, and indoor positioning \cite{bi_2023_psosvrpos,guo_2023_fedpos}. Since traditional GPS often underperforms indoors, there is significant demand for more efficient indoor localization techniques \cite{liu_2020_survey,hernndez_2021_wifinet}. As a core component of localization systems, precision position matching can reduce location errors in fingerprint-based algorithms \cite{bi_2023_psosvrpos, shang_2022_overview}. It also proves beneficial across multiple domains, including medical treatments \cite{blomberg_2018_responsibility}, smart venues \cite{kinoshita_2018_an}, and industrial Internet-of-Things \cite{ma_2020_binocular}. For example, precise medical scanning requires ensuring the patient is in a designated position \cite{blomberg_2018_responsibility}, and in smart homes, position matching can be used to verify resident presence \cite{tan_2020_iot}.

Position matching and estimation focus on different aspects of indoor positioning systems. Unlike localization estimation, which requires multiple reference points to predict an unknown position, position matching, commonly used for detecting static devices, determines whether the target is in a known position \cite{ma_2020_binocular}. The mediums to achieve it include cameras and Wi-Fi \cite{jang_2023_survey}. Compared to vision-based techniques, Wi-Fi-based algorithms offer unique strengths: lower risks of privacy infringement \cite{yang_2018_carefi}, flexibility in the unlicensed spectrum \cite{guo_2023_fedpos}, less need for additional equipment \cite{hernndez_2021_wifinet}, and adaptation under non-line-of-sight (NLOS) conditions \cite{khan_2023_crosscount}.

Furthermore, Wi-Fi-based localization methods have primarily utilized CSI and received signal strength indicator (RSSI) as key metrics \cite{shang_2022_overview}. CSI, which provides more granular information than RSSI, has spurred advancements in indoor localization by integrating machine learning algorithms \cite{kinoshita_2018_an} and physical similarity measurements \cite{yaro_2024_enhancing}. Compared to machine learning-integrated systems, which are often criticized for extensive training \cite{guo_2023_fedpos} and opaque decision-making processes \cite{bast_2020_mamimo}, physical index-based methods offer distinct advantages such as reduced complexity and greater transparency \cite{han_2015_cosine}. However, devising an effective physical measurement strategy is challenging due to CSI's variability and sensitivity to targets and its minor movements\cite{yang_2018_carefi}.

\begin{figure*} [ht]
    \centering
    \includegraphics[width=0.8\linewidth]{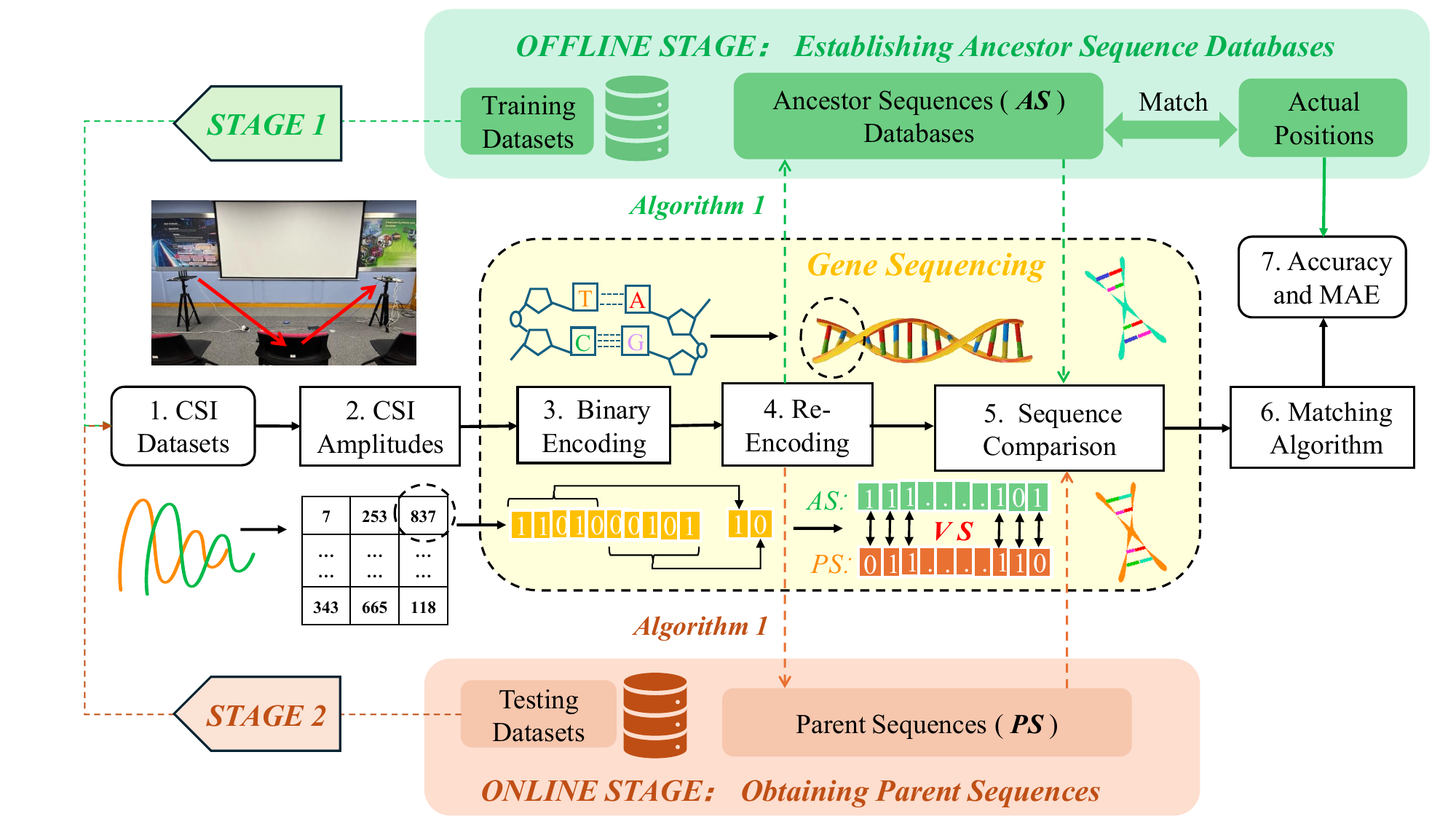} 
    \caption{The flowchart of the proposed algorithm.} 
    \label{fig:1} 
\end{figure*}

Inspired by gene sequencing techniques, this paper proposes a novel algorithm that combines binary encoding and fingerprinting, aiming to develop an effective physical measurement-based strategy for position matching. In gene sequencing, bases at corresponding positions tend to be identical among sequences that share close genetic relationships \cite{yang_2020_review}, with base similarity serving as the physical measure. Similarly, fingerprint-based localization leverages the principle that signal matrices from geographically proximate points often exhibit resemblances \cite{wang_2016_csibased}. Drawing on the similarities between the two methodologies, this work explores the conversion of each sample in CSI matrices into a comparable sequence.

While gene sequence comparison relies on the four inherent bases, the variability in the raw CSI matrix makes its direct comparisons impractical. Therefore, the proposed algorithm initially quantifies the raw CSI matrix into binary sequences, as illustrated in the yellow box in Fig.~\ref{fig:1}. The binary conversion enables the evaluation of sequence similarity by comparing bits at corresponding positions. Since Hamming distances represent the similarity between binary sequences, they are naturally utilized as the physical similarity measurement. The main contributions of this paper include:

\begin{itemize}
  \item \textbf{Novel Position Matching Algorithm:} Based on Wi-Fi and Raspberry Pi 4B, this paper develops an innovative algorithm that combines binary encoding with fingerprint-based techniques to enhance position matching. The results show that BiCSI achieves an average accuracy exceeding 98\% and a MAE of less than three centimeters across various scenarios.
    
  \item \textbf{Efficient Feature Vector Extraction:} This research introduces an advanced method for extracting feature vectors from CSI matrices that requires only kilobytes of storage, significantly less than the megabytes typically needed in traditional machine learning models.

  \item \textbf{Good Adaptability and Robustness:} The proposed algorithm is applicable to multiple similarity measurements, with the MAE of employing Hamming distance improving by 4\%-28\%. Additionally, it remains robust across different time periods, further enhancing its utility and versatility in diverse scenarios.

\end{itemize}

The rest of the paper is organized as follows. Section {\uppercase\expandafter{\romannumeral 2}} introduces the background and Section {\uppercase\expandafter{\romannumeral 3}} details the algorithm flowchart. Section {\uppercase\expandafter{\romannumeral 4}} discusses the experimental results. Finally, Section {\uppercase\expandafter{\romannumeral 5}} concludes the paper.

\section{Backgroud}
This section introduces the relevant techniques and definitions, including fingerprint-based techniques, RSSI and CSI, and classifications of different target states.

\subsection{Fingerprint-Based Localization}

Fingerprint-based methods have been extensively utilized for indoor positioning, leveraging fingerprints derived from various wireless signal indicators. Commonly used metrics include RSSI \cite{hernndez_2021_wifinet} and CSI \cite{liu_2023_towards}. Fingerprint-based systems collect signal metrics at reference locations and store them as feature vectors during the offline stage \cite{liu_2020_survey}. In the online stage, the system predicts the target location by comparing current signal metrics with the stored features \cite{tan_2020_iot}. These techniques are effective in mitigating the impact of direct signal interference \cite{shang_2022_overview}. When all potential positions are predefined and trained during the offline stage, fingerprint-based position estimation equals a position matching process.

Fingerprint-based techniques have primarily evolved by integrating machine learning algorithms \cite{kinoshita_2018_an} and using physical similarity measurements \cite{yaro_2024_enhancing}. Machine learning, with its dynamic capabilities, leverages methods like convolutional neural networks (CNN) \cite{bast_2020_mamimo}, genetic algorithms \cite{bi_2023_psosvrpos}, and support vector machines (SVM) \cite{zhang_2018_indoor}. However, challenges of these approaches include extensive data requirements \cite{guo_2023_fedpos} and opaque `black box' nature \cite{bast_2020_mamimo}. Therefore, physical metric-based methods are being developed due to their reduced complexity and greater transparency \cite{han_2015_cosine}. Using cosine similarity, Han et al. achieved a 20\% reduction in positioning errors under two meters \cite{han_2015_cosine}. Similarly, based on Euclidean distance, the weighted k-nearest neighbors (KNN) algorithm achieved a mean location error of less than one meter \cite{poulose_2020_performance}. However, the sensitivity of wireless metrics to targets, which leads to variability in the collected matrices, poses a challenge in devising an effective physical measurement strategy \cite{yang_2018_carefi}.

This paper introduces an innovative position matching algorithm based on binary encoding and fingerprint-based techniques. We convert CSI matrices into binary sequences and utilize Hamming distances to assess signal similarity. Unlike position estimation, which seeks more and closer reference points to enhance precision \cite{liu_2020_survey}, position matching benefits from a much sparser distribution of pre-trained points \cite{jang_2023_survey}. Consequently, our experiments are designed with testing positions spaced at meter-level intervals.

\subsection{RSSI and CSI}

RSSI and CSI have become popular signal metrics in fingerprint-based techniques. RSSI is a MAC-layer parameter, while CSI is a physical-layer parameter \cite{shang_2022_overview}. RSSI, which correlates with the distance between transceivers, is modeled using various approaches, such as the free space model \cite{mollel_2014_an} in wireless networks. However, RSSI alone has proven less reliable for indoor positioning due to the presence of rich multipath effects \cite{khan_2023_crosscount}. Consequently, RSSI-based methods have often been supplemented with machine learning algorithms. For instance, the CNN model in \cite{xiong_2023_highprecision} improved the location accuracy by 6\% than traditional methods. Integrating the support vector regression with the particle swarm optimization (PSOSVR) further achieved an average positioning error of around one meter \cite{bi_2023_psosvrpos}. By contrast, CSI offers more fine-grained information, thus potentially enhancing performance in indoor localization scenarios \cite{wang_2016_csibased}.

\label{deqn_ex1a}
\begin{equation}
    \mathbf{C}^{(p,q)} = \sum_{p=1}^{P} \sum_{q=1}^{Q} \left( |C_{i,k}^{(p,q)}| e^{j\theta_{i,k}^{(p,q)}} \right)
\end{equation}

\begin{equation}
C_{i,k} = \begin{pmatrix}
|C_{1,1}| e^{j\theta_{1,1}} & \cdots & |C_{1,k}| e^{j\theta_{1,k}} \\
\vdots & \ddots & \vdots \\
|C_{i,1}| e^{j\theta_{i,1}} & \cdots & |C_{i,k}| e^{j\theta_{i,k}}
\end{pmatrix}
\end{equation}

According to \cite{guo_2023_fedpos,khan_2023_crosscount, bast_2020_mamimo}, CSI can be represented in (1) and (2). $\mathbf{C}^{(p,q)}$ represents that CSI is related to \textit{P}
 (the number of transmitters or their antennas) and \textit{Q} (the number of receivers or their antennas). For each pair of transceivers (or antennas) in (2), $|C_{i,k}|$ represents the amplitude of the $i$-th sample and the $k$-th subcarrier; $\theta_{i,k}$ represents the phase of the $i$-th data sample and the $k$-th subcarrier.

As demonstrated in (1), CSI contains amplitude and phase information. Due to higher noise levels in the phase matrix \cite{wang_2016_csibased}, the amplitude matrix has gained more popularity in CSI-based studies. Utilizing CSI amplitudes, the federated learning approach in \cite{guo_2023_fedpos} achieved an average localization error of under 0.5 meters. Similarly, the neural network in \cite{gonultas_2022_csibased} reduced the median position error to the centimeter level for static devices. Additionally, common devices for acquiring CSI from Wi-Fi include Hamming Board \cite{khan_2023_crosscount} and Raspberry Pi 4B \cite{guo_2023_fedpos}, with the latter capable of extracting more subcarriers. Consequently, this paper utilizes the Raspberry Pi 4B to acquire CSI amplitudes as the metric.

\subsection{Definitions of Target States}

Position matching has been widely developed for detecting static devices, such as indoor positioning of equipment \cite{guo_2023_fedpos}, and target grasping of robots \cite{ma_2020_binocular}. However, one limitation of existing research has been less focus on human targets than on static devices \cite{shang_2022_overview}, partly due to body movements' impact on system accuracy. Authors in \cite{ding_2024_robust} found that body movements can cause continuous fluctuations in CSI matrices, challenging the effectiveness of algorithms. Furthermore, the impact varies when targets are under different states \cite{yang_2018_carefi}. Consequently, understanding the state of targets is crucial for improving radio-based localization systems.

This section introduces the distinctions among dynamic, stationary, and semi-stationary states for Wi-Fi-based indoor positioning systems. The dynamic state involves activities that cause significant positional changes, such as walking \cite{qian2018widar2}. The stationary state, in contrast, is marked by a complete absence of movement, typical for most devices and non-living entities \cite{belalkorany_2021_counting}. However, this binary classification fails to capture a state inherent to most living beings; they exhibit subtle, unavoidable movements like breathing and heartbeat even when seemingly stationary. This intermediate state, termed `semi-stationary' \cite{khan_2023_crosscount}, is exemplified by a person sitting; while largely stable, minor movements such as fidgeting may still occur \cite{jiang_2024_pacount}.

Although prevalent in daily activities, the semi-stationary state has received less attention in existing research \cite{belalkorany_2021_counting}. This paper aims to alleviate the more pronounced CSI fluctuations in semi-stationary targets compared to static ones through the proposed binary encoding steps, thereby expanding the applicability of position matching systems and enhancing the effectiveness of localization-based sensing services.

\section{THE PROPOSED ALGORITHM}

The proposed BiCSI algorithm combines binary encoding with fingerprint-based techniques. As depicted in Fig.\ref{fig:1}, the algorithm's flowchart consists of seven steps across two stages. \textbf{Step 1} involves data collection using Raspberry Pi 4B to gather CSI data from Wi-Fi. \textbf{Step 2} extracts the integer part of CSI amplitudes \((AP)\). The resulting amplitude matrix is arranged like (2), with packets listed by rows and subcarriers by columns. The subsequent three core steps, highlighted in the yellow box in Fig.\ref{fig:1}, simulate gene sequencing processes: binary encoding, re-encoding, and sequence comparison.

\begin{figure} [ht]
    \centering
    \includegraphics[width=0.9\linewidth]{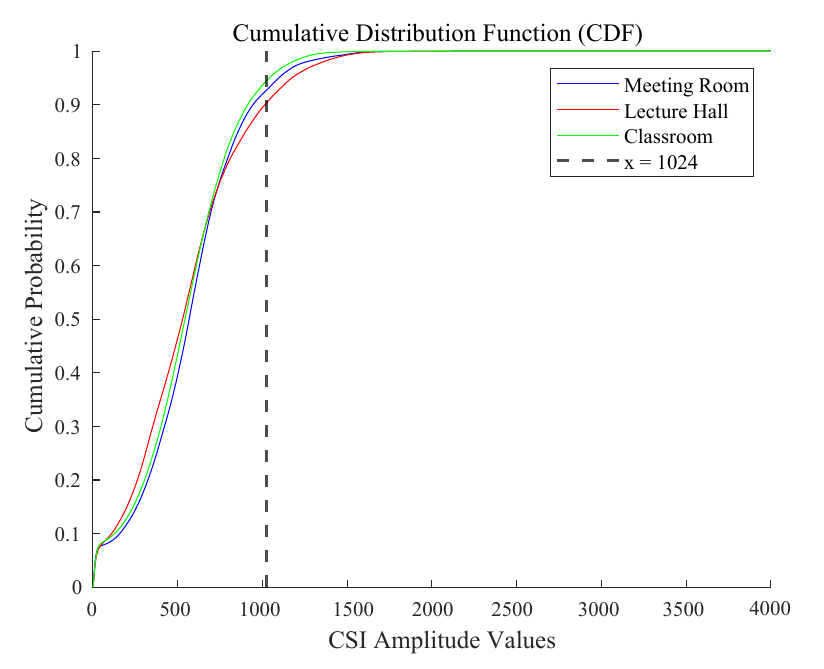}
    \caption{The CDF of CSI amplitudes.}
    \label{fig:2}
\end{figure}

\textbf{Step 3} converts CSI matrices to binary sequences through the binary encoder in (3). It converts the integer amplitude to a ten-bit binary sequence \((B_1-B_{10})\). Specifically, values less than 1024 are converted into the corresponding binary bits; values greater than 1024 are represented by the binary bits of 0. Selecting ten bits (representing 1023 maximum) is based on the CSI amplitude distribution. As illustrated in Fig.~\ref{fig:2}, over 90\% of CSI amplitudes fall below 1024. Therefore, this binary encoder keeps most amplitudes, balancing the retained information and the number of bits occupied.

\begin{equation}
    \mathbb{B} = \begin{cases} 
    B_{10}(\text{AP}), & \text{if } \text{AP} < 1024 \\
    B_{10}(0), & \text{if } \text{AP} \geq 1024
    \end{cases} 
\end{equation}

\noindent where \( B_{10}(x) \) represents a binary encoder with an output of ten bits.

\begin{equation}
    \begin{cases} 
    H = u\left(\sum_{j=1}^5 \mathbb{B}_j - 3\right) \\
    L = u\left(\sum_{j=6}^{10} \mathbb{B}_j - 3\right) 
    \end{cases} 
\end{equation}

\noindent where $u(x)$ is a condition function applied to the j-th bits, equal to 1 if $x \geq 0$, and 0 otherwise.

\begin{equation}
    NB_{i,k} = H_{i,k} L_{i,k}, \quad H_{i,k}, L_{i,k} \in \{0, 1\}
\end{equation}

\begin{equation}
C_{i,k} = \begin{pmatrix}
GS_1 \\
\vdots \\
GS_i
\end{pmatrix}
= \begin{pmatrix}
NB_{1,1} & \cdots & NB_{1,k}  \\
\vdots & \ddots & \vdots \\
NB_{i,1}  & \cdots & NB_{i,k} 
\end{pmatrix}
\end{equation}

\noindent where i and k represent the i-th sample and the k-th subcarriers in the CSI matrix.

\begin{equation}
D(AS, PS) = \sum_{j=1}^n \delta(AS, PS)
\end{equation}

\noindent where \( \delta(x)\) is an indicator function. Its result is 1 when the j-th bit of two gene sequences, \( AS_j \) and \( PS_j \), are different; otherwise, it is 0. \( D(AS, PS) \) represents the Hamming distance between sequences \( AS \) and \( PS \).

\textbf{Step 4} is the re-encoding step. It applies (4) to re-encode each ten-bit binary into a new two-bit binary (\(NB\)). Initially, it equally divides the ten bits into high-order and low-order parts, each containing five bits. The re-encoding of five bits is designed to mitigate signal fluctuations caused by body movements. Due to the sensitivity of CSI, even slight movements can alter the signal matrix \cite{yang_2018_carefi,ding_2024_robust}. The re-encoding step alleviates this issue by replacing each segment with the more frequently occurring bit value. For example, if the lower-order part of \( L_{1,1} \) is 00000 and the lower part of \( L_{2,1} \) is 00101, then their L values obtained after applying (4) are both 0, which helps filter out some CSI fluctuations.

Moreover, since high-order bits represent larger decimal values than the low-order, this division helps preserve the original amplitude information. After performing modulo two additions on each of five bits, the high-order parts are encoded as \(H\), and the low-order parts as \(L\), represented by (5). As shown in Fig. 1, each decimal CSI value is finally represented as a two-bit binary after the re-encoding. Then, the CSI matrix in (2) is converted into (6), and each row is termed a gene sequence (\(GS\)) in this work. 

\textbf{Step 5} refers to the sequence comparison step. The similarity between sequences is assessed by verifying if bits in corresponding positions are identical, which aligns with the definition of Hamming distance in the communications field. BiCSI then applies the Hamming distance as the similarity measurement in (7). Specifically, there are two types of gene sequences. As shown in Fig.~\ref{fig:1}, the feature vectors obtained in the offline stage are ancestor sequences (\(AS\)), while the binary sequences in the online stage are parent sequences (\(PS\)).

\textbf{Step 6} is the matching algorithm utilizing the Hamming distances from the last step. For each parent sequence, the ancestor sequence with the minimum Hamming distance is selected, and the corresponding position of the ancestor sequence is outputted as the predicted position.

\textbf{Step 7} calculates the error using two performance indicators: the MAE and accuracy. The calculation method for MAE is given in (8), while accuracy is obtained in (9). In the two equations, \(n\) is the number of data points. Position matching is used to determine whether the target is located at a known position by comparing its predicted coordinates, \(P(x, y)\), with the coordinates of reference points, \(A(x, y)\). \(| P (x, y) - A (x, y) |\) then represents the absolute sum of the differences between the horizontal and vertical coordinates of each predicted point and the actual point. Considering horizontal and vertical aspects, \(2n\) is divided by the mean absolute location error.

\begin{equation}
    \text{MAE} = \frac{\sum_{i=0}^{n} \left| P_i(x,y) - A_i(x,y) \right|}{2n} 
\end{equation}

\begin{equation}
    \text{Accuracy} = \frac{\text{The number of correct predictions}}{n} 
\end{equation}

The two stages are based on Steps 1-4: offline and online, as illustrated in Fig.~\ref{fig:1}. \textbf{Stage 1}, similar to the offline stage in fingerprint-based localization, collects signal features of each position and stores them in an offline database. As shown in Algorithm 1, which organizes binary sequences into rows (i-th) and bits into columns (j-th), the extracted method derives the ancestor sequences based on the binary sequences from Step 4. The bits in each column of the ancestor sequence are determined by the corresponding columns in the training samples, which consist of 12,000 packets. Owing to the volatility of the CSI, each column in binary sequences collected from the same location typically exhibits two potential values, 0 or 1, leading to two possible ancestor sequences (\(AS^1\) and \(AS^2\)) for each position. However, as mentioned in Step 4, CSI matrices are prone to fluctuations due to body movements, challenging the effectiveness of feature vectors. To further mitigate the adverse effects of such fluctuations, Algorithm 1 calculates the frequencies of 0s (\(N_0\)), 1s (\(N_1\)), and the gap between them for each column.

Specifically, two scenarios are considered based on the gap and a predefined threshold (\(Tr\)). First, if the gap exceeds \(Tr\), indicating a significant dominance of one value over the other, the assignment is as follows: if \(N_0 > N_1\), both ancestor sequences are assigned a `0' for that column; otherwise, they are assigned a `1'. Second, if the gap is less than \(Tr\), suggesting a balanced presence of both zeros and ones, BiCSI retains both bits: it sets the column in the first sequence to `1' and in the second sequence to `0'.

Finally, these ancestor sequences obtained by Algorithm 1 are correlated with corresponding positions and stored in the offline database as feature vectors. Due to the re-encoding, the storage space required for these feature vectors is reduced by 80\% compared to the storage of original CSI matrices, which use a decimal format (ten bits per value) in Step 3.

\begin{algorithm}[ht]
\caption{Getting Ancestor Sequences}\label{alg:alg1}
\begin{algorithmic}
\STATE \textbf{Input:} Gene Sequences ($GS_i$)
\STATE Set a threshold ($Tr$)
\FOR{each column ($j$)}
    \STATE Count the number of 0 ($N_0$) in $GS_i$
    \STATE Count the number of 1 ($N_1$) in $GS_i$
    \IF {$|N_0 - N_1| \geq Tr$}
        \IF{$N_0 > N_1$}
            \STATE {$AS_j^1$ = 0} and {$AS_j^2$ = 0}
        \ELSE
            \STATE {$AS_j^1$ = 1} and {$AS_j^2$ = 1}
        \ENDIF
    \ELSE
        \STATE {$AS_j^1$ = 1}
        \STATE {$AS_j^2$ = 0}
    \ENDIF
\ENDFOR
\STATE \textbf{Output:} Two Ancestor Sequences ({{$AS^1$} and {$AS^2$}})\\
\end{algorithmic}
\end{algorithm}

\textbf{Stage 2} is the online stage, where parent sequences are obtained for testing. During this phase, Algorithm 1 functions similarly to the offline stage but with two differences. The first difference concerns the volume of data handled. Given the limited computing capabilities of edge devices, one parent sequence is derived from every 120 packets based on our experimental observations. The second difference involves the method of bit setting, which is determined by the frequency of 1s and 0s without applying a threshold due to fewer samples. Then, as mentioned in Steps 5-7, the subsequent steps calculate Hamming distances, predict positions, and compute the MAE and accuracy.

\begin{figure*}[ht]  
    \centering
    \begin{subfigure}[b]{0.5\textwidth}  
        \centering
        \includegraphics[width=0.9\linewidth]{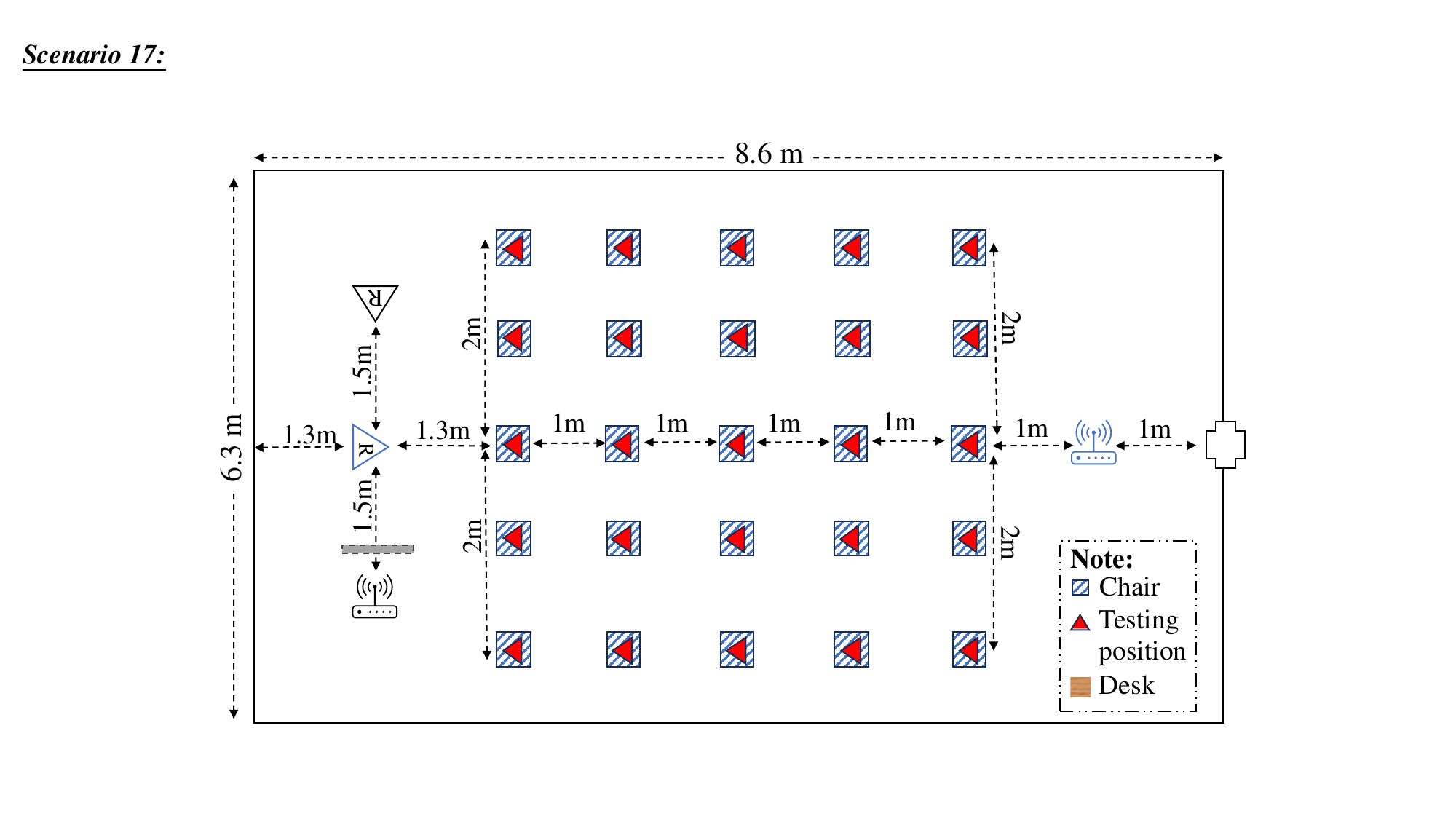}
        \caption{}
        \label{fig:3a}
    \end{subfigure}%
    \begin{subfigure}[b]{0.5\textwidth}
        \centering
        \includegraphics[width=0.9\linewidth]{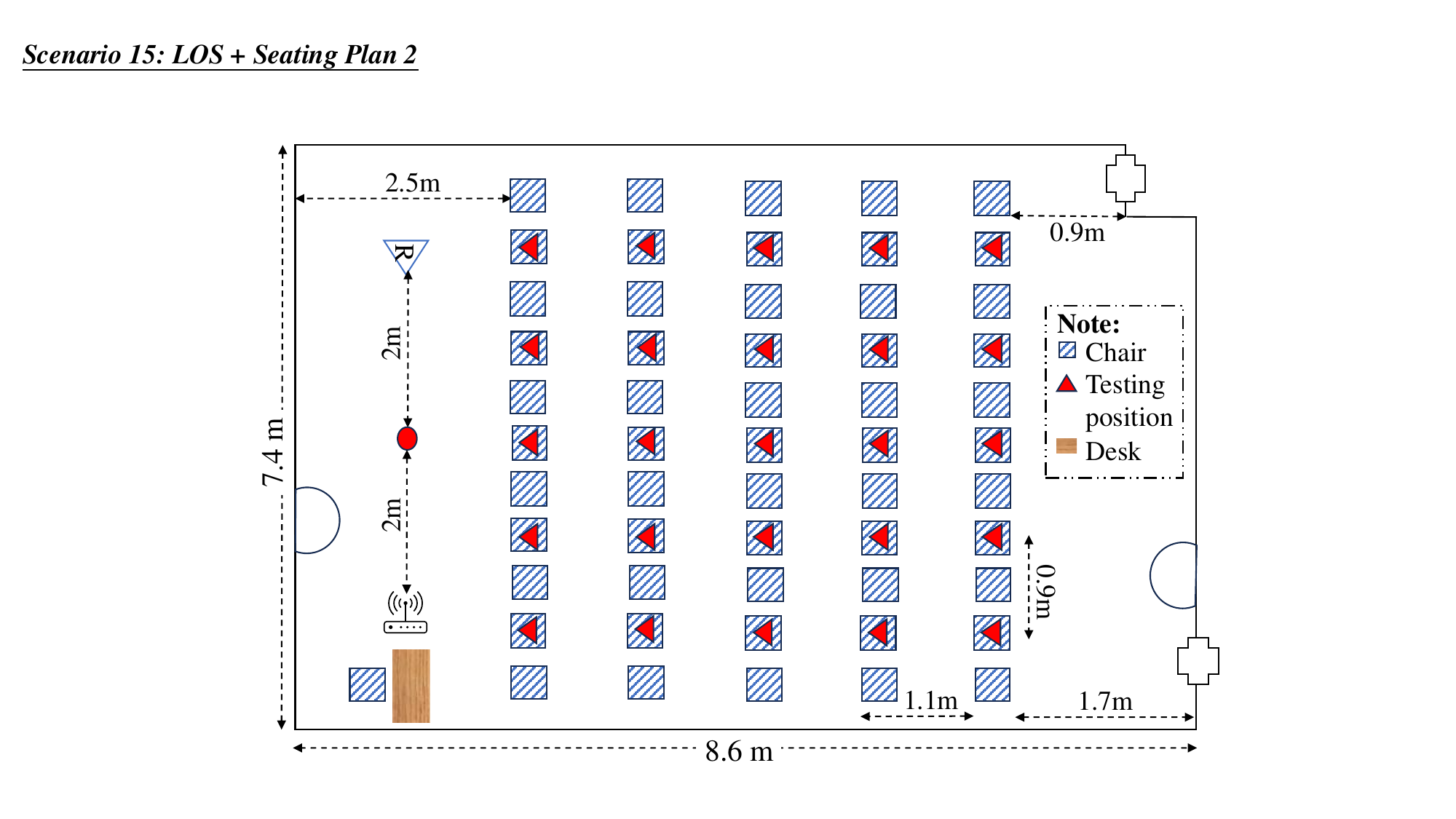}
        \caption{}
        \label{fig:3b}
    \end{subfigure}
    
    \begin{subfigure}[b]{0.5\textwidth}
        \centering
        \includegraphics[width=0.9\linewidth]{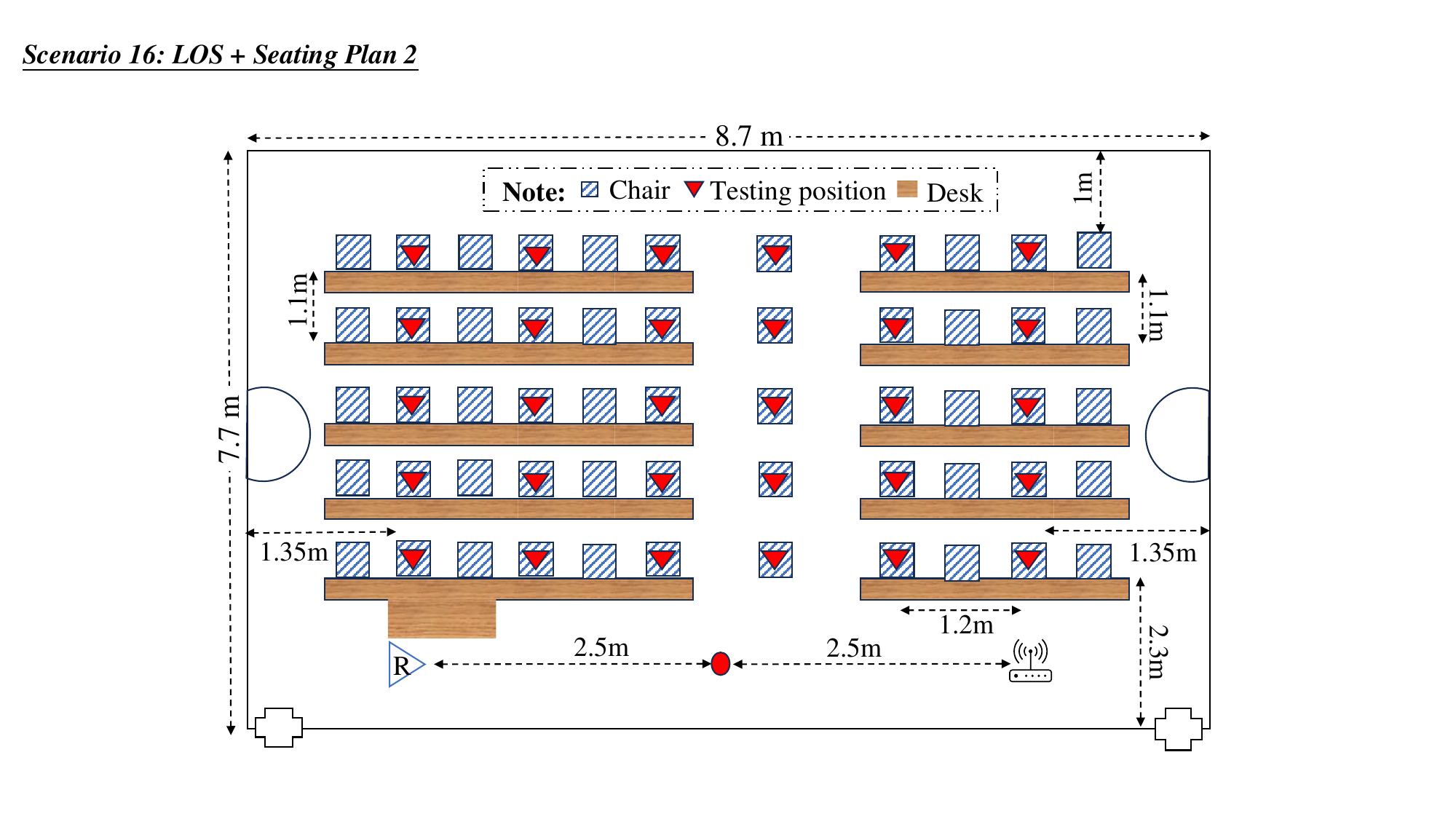}
        \caption{}
        \label{fig:3c}
    \end{subfigure}%
    \begin{subfigure}[b]{0.5\textwidth}
        \centering
        \includegraphics[width=0.75\linewidth]{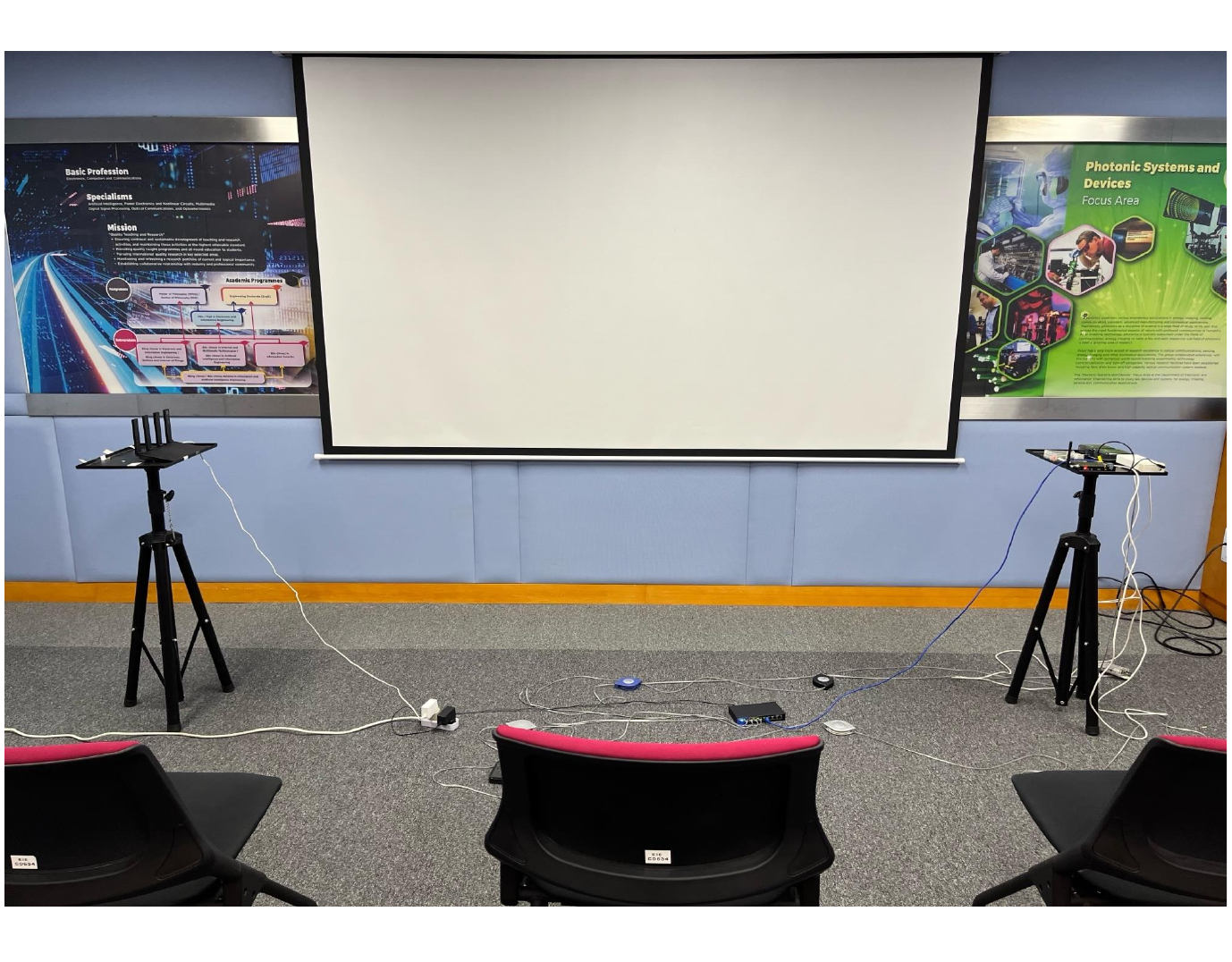}
        \caption{}
        \label{fig:3d}
    \end{subfigure}
    \caption{Five scenarios. (a) Meeting room with Scenario A1-A3. (b) Scenario B: Lecture hall. (c) Scenario C: Classroom. (d) Experiment setup.}
    \label{fig:3}
\end{figure*}

\section{EXPERIMENTAL RESULTS AND EVALUATION}
This section evaluates the algorithm's performance across various testing scenarios, assessing the impact of different parameters, including physical measurements, threshold settings, temporal factors, and transceiver distances.
\subsection{Experiment Setup}

As depicted in Fig.~\ref{fig:3}, the experiments encompass five testing scenarios, exploring the algorithm's performance under different spatial arrangements. Fig.~\ref{fig:3a} illustrates the primary testing environment within a meeting room, hosting three scenarios. In Scenarios A1 and A2, the transceivers were positioned on the left side of the room (Tp1: marked in black). Scenario A1 was set up for the LOS, and Scenario A2 for the NLOS with a 2 m $\times$ 1.5 m barrier. In Scenario A3, transceivers were located at opposing ends of the room (Tp2: marked in blue) with the LOS. Fig.~\ref{fig:3b} details Scenario B, and Fig.~\ref{fig:3c} presents Scenario C.

Fig.~\ref{fig:3d} further illustrates our experimental setup and Table~\ref{tab: experiment configurations} lists the detailed configurations utilized in this work. Based on Raspberry Pi 4B, our system can extract 256 CSI subcarriers from a commercial Wi-Fi that operates at 5 GHz with an 80 MHz bandwidth. Transceivers were positioned at a height of 95 cm, and Raspberry Pi 4B was configured to pair with an access point and connected to a device with the Nexmon \cite{guo_2023_fedpos}. All devices were interconnected through a network switch. Additionally, this work excluded 26 subcarriers, including the pilot referenced in \cite{guo_2023_fedpos} and seven invariant subcarriers. A laptop was used to manage data transmission and collection, and data processing was performed using Matlab R2023a.

\begin{table}[ht]
  \centering
  \caption{Experiment Configurations}
  \label{tab: experiment configurations}
  \begin{tabular}{@{}ll@{}}
    \toprule
    \multicolumn{2}{c}{Configuration} \\
    \midrule
    Transceiver & Huawei Router \\
    Receiver (R) & Raspberry Pi 4B \\
    Frequency Band & 5 GHz \\
    Packet Rate & 1000 Hz \\
    Number of sub-carriers & 256 (230 used) \\
    Training samples & 12000 packets \\
    Testing samples & 24000 packets \\
    \bottomrule
  \end{tabular}
\end{table}

Moreover, each scenario was tested six times, and average values were calculated to minimize random errors. In each test, packets were collected at 1000 Hz. The data from each test was stored in separate files, facilitating independent separation into offline and online stages. For each testing point, one-third of the datasets were used for the offline stage, while the remaining two-thirds were utilized for testing. The CSI data for every testing position was collected continuously to minimize variability caused by environmental changes, such as chair movements. Additionally, the threshold (\(Tr\)) employed in Algorithm 1 was set at 5\% of the training dataset size.

\subsection{Algorithm Performance in Different Scenarios}

This work explores the performance of different algorithms on the CSI dataset. Specifically, three categories of fingerprint-based positioning algorithms are selected for evaluation. The first category includes matching methods utilizing physical similarity measurements such as Hamming distances used in our proposed algorithm (BiCSI), cosine similarity \cite{han_2015_cosine}, and Pearson correlation coefficient \cite{yang_2018_carefi}. The cosine similarity and Pearson correlation coefficient rely on the raw CSI data. The second category comprises traditional machine learning algorithms, including the CNN \cite{xiong_2023_highprecision}, KNN \cite{poulose_2020_performance} and SVM \cite{zhang_2018_indoor}. In particular, KNN, which uses the Euclidean distance, can be considered a hybrid of physical measurements and machine learning approaches. The third category includes novel fingerprint-based positioning techniques, such as PSOSVR \cite{bi_2023_psosvrpos} and FedPos \cite{guo_2023_fedpos}. Since location estimation equates to position matching for all trained test points, MAE can effectively serve as a performance metric to evaluate these algorithms. 

\begin{figure}[ht] 
    \centering
    \includegraphics[width=0.9\linewidth]{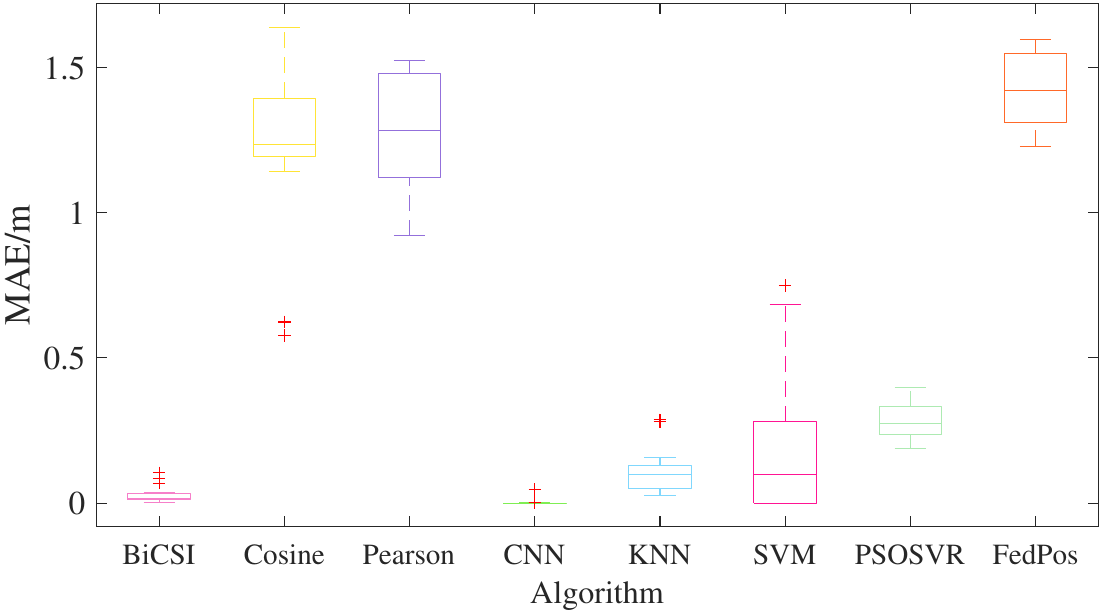}
    \caption{The MAE of each algorithm across scenarios.}
    \label{fig: MAE1}
\end{figure}

The MAE results are illustrated in Fig.~\ref{fig: MAE1}. Firstly, the proposed algorithm performs comparably to traditional machine learning algorithms across various scenarios, exhibiting lower average MAE values than other categories. Specifically, CNN and BiCSI demonstrate the best performance, with their average MAEs less than three centimeters, showcasing exceptional location matching capabilities. The variation range for BiCSI is less than 0.15 m, following CNN which is less than 0.1 m. The average MAEs of SVM and KNN are very close across different scenarios, yet the variance of SVM is more than twice that of KNN. Secondly, BiCSI outweighs other similarity measurements-based methods. The average MAEs (over 1.2 m) and the variation ranges of cosine similarity and Pearson correlation coefficient are significantly greater than BiCSI's. Algorithms in the third category exhibit markedly different performances, largely due to their designs for specific scenarios \cite{bi_2023_psosvrpos,guo_2023_fedpos}; moreover, limited applicability is a common issue with machine learning-based algorithms \cite{liu_2020_survey}.

\begin{figure}[ht] 
    \centering
    \includegraphics[width=0.8\linewidth]{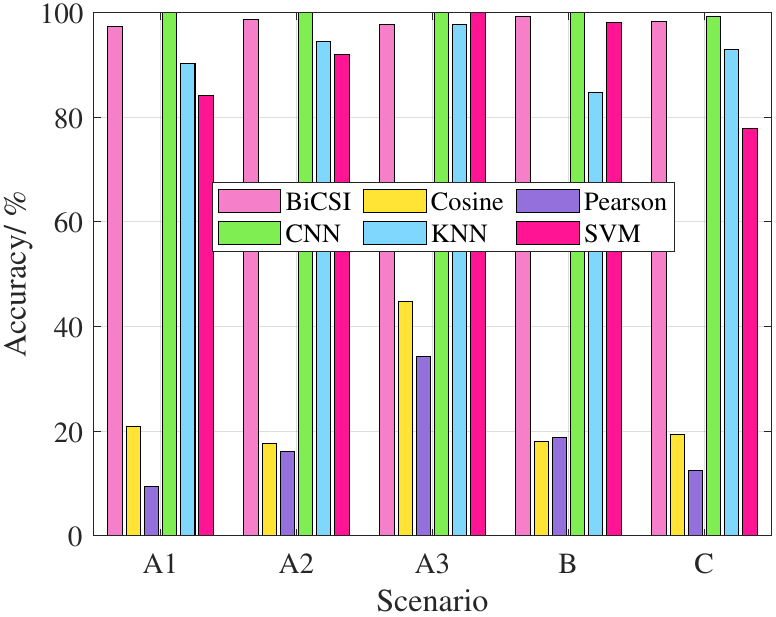}
    \caption{The accuracy of each algorithm on each scenario.}
    \label{fig: Accuracy1}
\end{figure}

Fig.~\ref{fig: Accuracy1} displays the accuracy matrix of each algorithm in different scenarios. PSOSVR and FedPos, explicitly designed for position estimation, are not evaluated for matching accuracy. The performance of most algorithms remains stable from Scenario A1 to A2, indicating robust detection capabilities even when the direct path is obstructed. Additionally, accuracy improves significantly in Scenario A3 compared to A1, attributed to more seats falling within smaller Fresnel zones when the transmitter and receiver are positioned at Tp2 \cite{liu_2023_towards}, enhancing detection capabilities.

Moreover, BiCSI achieves a remarkable average accuracy above 98\% across tests. While its accuracy is slightly lower than that of CNN, BiCSI significantly outperforms algorithms based directly on similarity measurements by at least two-fold. Despite high accuracy, as detailed in Table~\ref{tab:training_times}, CNN and BiCSI require longer processing times than KNN and SVM. Based on the variation range of training time across scenarios, SVM fluctuates most, which is similar to its MAE and accuracy metrics. The latency in BiCSI is attributed to its two-step process of binary encoding and feature extraction, while other algorithms process raw CSI data directly; meanwhile, the latency of CNN is related to the size of the datasets. Despite the additional binary encoding step, BiCSI remains about 26\% faster than CNN, effectively managing complex tasks while maintaining high accuracy.

\begin{table}[htbp]
\centering
\caption{The Dataset Training/Processing Times of Algorithms across Scenarios.}
\label{tab:training_times}
\begin{tabular}{lcccc}
\hline
\textbf{Methods} & \textbf{BiCSI} & \textbf{CNN\cite{xiong_2023_highprecision}} & \textbf{KNN\cite{poulose_2020_performance}} & \textbf{SVM\cite{zhang_2018_indoor}} \\
\textbf{Average Time/s}    & 1436           & 1942         & 514          & 729          \\
\textbf{Variation Range/s}  & 224            & 242          & 26           & 504          \\
\hline
\end{tabular}
\end{table}

Additionally, BiCSI significantly reduces storage space compared to traditional machine learning algorithms. The size of the trained model, which stores feature vectors, is obtained by the `whos' function in  MATLAB. Listed in Table~\ref{tab:model-sizes}, the storage space required by BiCSI is only in the kilobyte range, significantly less than that required by machine learning algorithms. The storage efficiency is achieved through three aspects. First, the binary encoding step allows feature vectors to be compactly stored as bits. Second, the re-encoding step further reduces the storage space required. Third, the process of deriving feature vectors in Algorithm 1 also contributes to additional storage savings. Overall, despite the latency of binary encoding steps, BiCSI achieves remarkable performances, maintaining an average MAE of less than three centimeters and exceeding an average accuracy of 98\%, while significantly reducing storage requirements. These attributes make BiCSI effective for detecting semi-stationary targets.

\begin{table}[ht]
\centering
\caption{The Storage Size of Trained Model in Algorithms}
\label{tab:model-sizes}
\begin{tabular}{lcccc}
\hline
\multirow{2}{*}{\textbf{Scenario}} & \multicolumn{4}{c}{\textbf{Storage Size}} \\ \cline{2-5}
                                   & \textbf{BiCSI} & \textbf{CNN\cite{xiong_2023_highprecision}} & \textbf{KNN\cite{poulose_2020_performance}} & \textbf{SVM\cite{zhang_2018_indoor}} \\ \hline
A1: LOS \& T1                      & 2.80 KB        & 0.21 MB      & 177.11 MB    & 176.63 MB    \\
A2: NLOS \& T1                     & 2.80 KB        & 0.21 MB      & 177.11 MB    & 176.59 MB    \\
A3: LOS \& T2                      & 2.80 KB        & 0.21 MB      & 177.11 MB    & 176.63 MB    \\
B                                  & 2.80 KB        & 0.21 MB      & 177.11 MB    & 176.63 MB    \\
C                                  & 3.37 KB        & 0.22 MB      & 212.53 MB    & 211.88 MB    \\ \hline
\end{tabular}
\end{table}

\subsection{Impact of Different Similarity Measurements on the Algorithm}

\begin{figure}[ht] 
    \centering
    \includegraphics[width=0.9\linewidth]{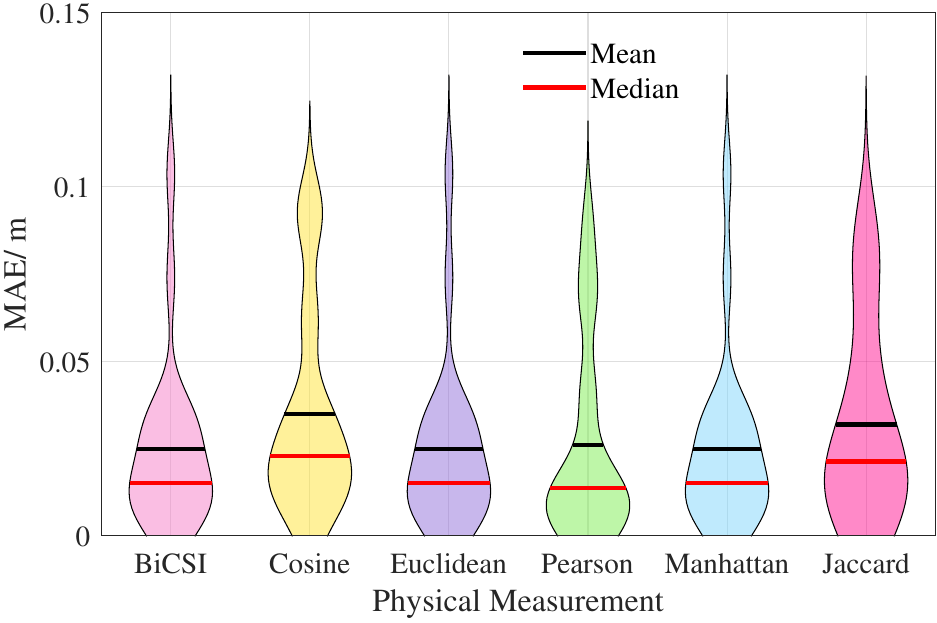}
    \caption{The MAE of each algorithm across scenarios.}
    \label{fig: MAE2}
\end{figure}

\begin{figure}[ht] 
    \centering
    \includegraphics[width=0.9\linewidth]{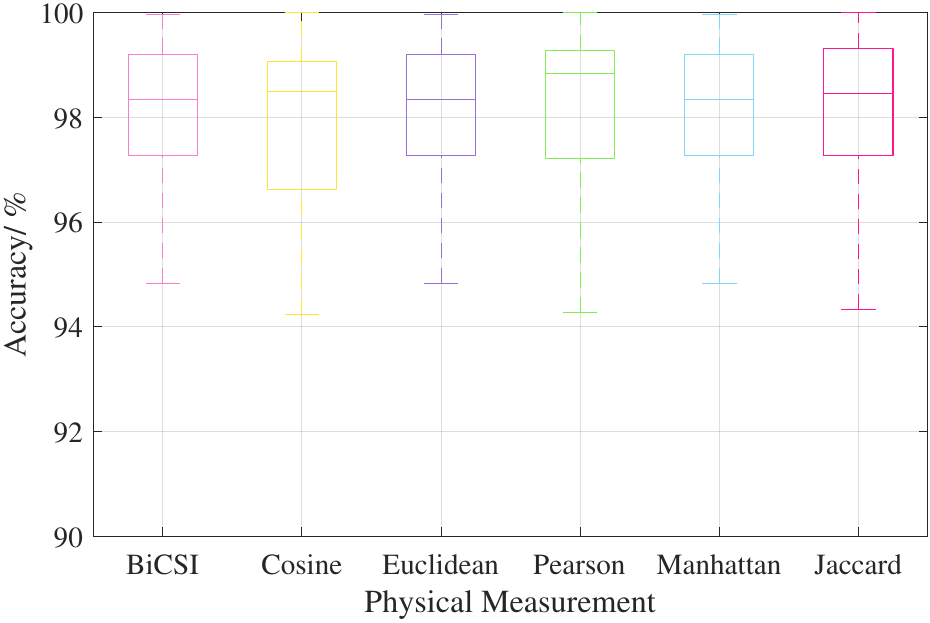}
    \caption{The accuracy of each algorithm across scenarios.}
    \label{fig: Accuracy2}
\end{figure}

In the above section, physical metrics underperformed on the raw CSI dataset compared to the proposed algorithm. However, it remains to be seen whether the superior matching capability stems from the specific similarity measurement (the Hamming distance) or the proposed binary encoding steps. To further validate that the designed encoding method effectively extracts feature vectors within the CSI matrix, this section introduces alternative physical metrics as replacements for the Hamming distance used on BiCSI for comparison. The selected similarity measurements are categorized into two types \cite{yaro_2024_enhancing}: distance metrics, which include the Euclidean distance and Manhattan distance; and correlation metrics, which include the cosine similarity, Pearson correlation coefficient, and Jaccard index. The results are shown in Fig.~\ref{fig: MAE2} (where the width indicates density) and Fig.~\ref{fig: Accuracy2}.

Despite utilizing different physical metrics, the algorithms' performance remains remarkably consistent across scenarios. Their average MAEs are less than 0.04 m (Fig.~\ref{fig: MAE2}), and the average accuracy exceeds 98\% (Fig.~\ref{fig: Accuracy2}). One reason for their close performance is that the same binary processes are applied, ensuring that the sequence compared across different physical indices remains consistent. Another reason is the inherent limited variability in binary sequences, which contain only two values: 0 and 1. The consistently superior performance across various similarity metrics indicates that the proposed binary encoding method effectively extracts features from CSI data. It also enables the algorithm to mitigate the reliance on a specific similarity measurement, enhancing its utility and versatility in different scenarios.

\begin{figure}[ht] 
    \centering
    \includegraphics[width=0.7\linewidth]{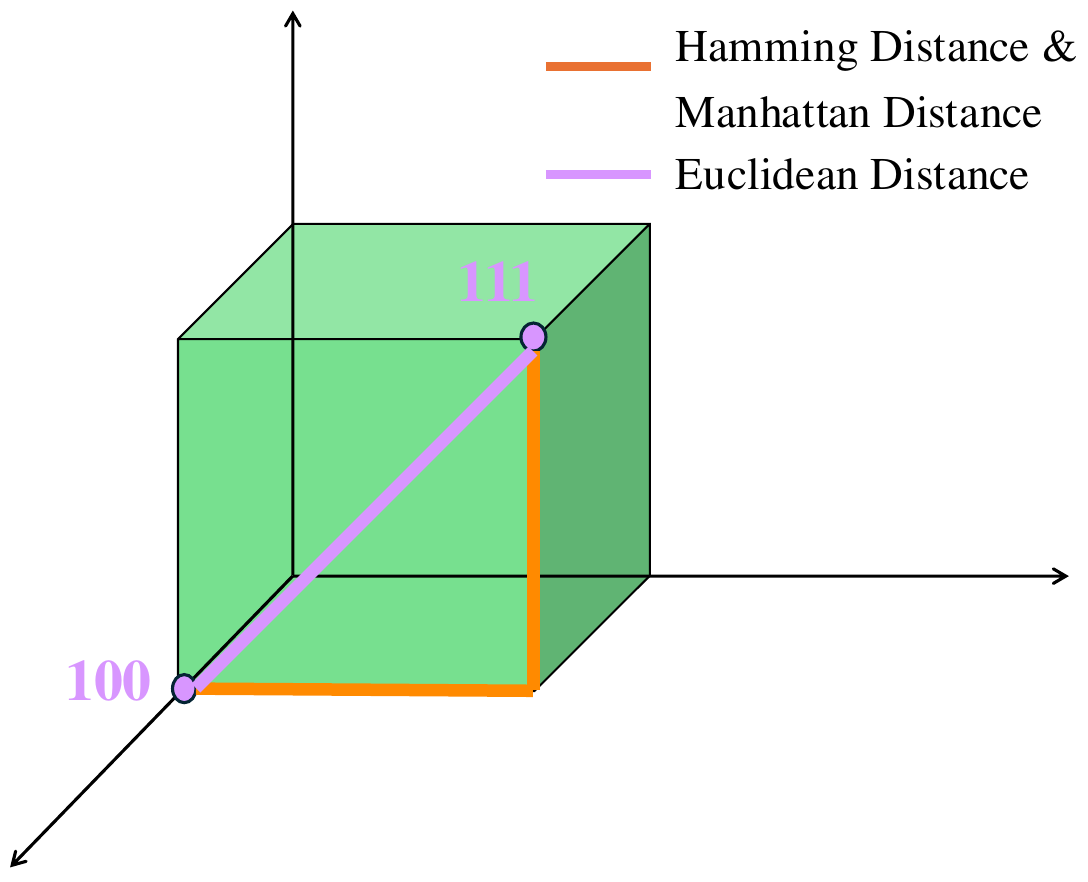}
    \caption{An example of the three types of distances on binary sequences.}
    \label{fig: Distances}
\end{figure}

Specifically, the performance of distance-based physical metrics is identical. The example in Fig.~\ref{fig: Distances} demonstrates that for binary sequences, the Hamming distance equates to the Manhattan distance and equals the square of the Euclidean distance. Since these three types of distance metrics are directly related to the Hamming distance, they exhibit the same performance across scenarios. Compared to correlation coefficients, Fig.~\ref{fig: Accuracy2} shows that the average accuracy achieved with these distance-based metrics is slightly lower.

However, utilizing Hamming distances offers three slight advantages. Firstly, as shown in Fig.~\ref{fig: Accuracy2}, the variation range in accuracy among distance metrics is smaller than correlation metrics, indicating their greater robustness. The second advantage pertains to the MAE. Fig.~\ref{fig: MAE2} shows that the proportion of MAE values exceeding 0.04 meters is smaller for distance metrics than correlation metrics, indicating that location errors above 0.04 meters occur less frequently with distance metrics. Additionally, the average MAE when employing the Hamming distance outperforms other metrics: it is 28\% better than cosine similarity, 4\% better than the Pearson correlation coefficient, and 21\% better than the Jaccard index.

\subsection{The Impact of Different Thresholds on the Algorithm}

This paper proposes an algorithm (Algorithm 1) for deriving ancestor sequences (feature vectors), where the threshold is a critical variable factor. The threshold value significantly influences the construction of feature vectors and, consequently, the algorithm's performance. For binary sequences, the matching accuracy is closely linked to the Hamming distance between sequences. A smaller Hamming distance implies reduced exclusivity among feature vectors, which could lead to higher matching errors. Therefore, we can efficiently assess the impact of different thresholds by the changing trend of Hamming distances between feature vectors.

\begin{figure}[ht] 
    \centering
    \includegraphics[width=0.9\linewidth]{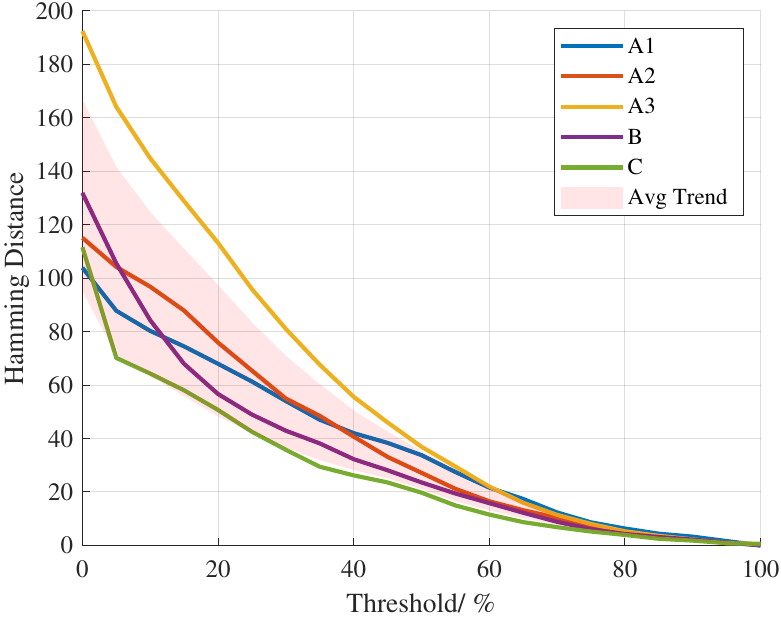}
    \caption{The average Hamming distance under different thresholds of each scenario.} 
    \label{fig: Hamming2}
\end{figure}

Fig.~\ref{fig: Hamming2} shows the changing trends of the average Hamming distance of each scenario under different thresholds. The threshold is expressed as a percentage of the training dataset size. As the threshold increases, the average Hamming distances tend to decrease, indicating a decline in algorithm performance. It occurs because as the threshold increases, Algorithm 1 is more inclined to execute the case where the gap is less than \(Tr\), causing the ancestor sequences obtained at different positions to become increasingly similar. Finally, when the threshold equals the size of the training dataset, the feature vectors obtained at each position become identical. As a result, the average Hamming distances reach zero, implying a total matching failure. Conversely, when the threshold is zero, Algorithm 1 only executes cases where the gap is greater than \(Tr\) (assigning the value with the most frequent bits). Based on experimental observations, this work sets the threshold at 5\% of the size of the training dataset.

\subsection{Algorithm Robustness Against Time}

This section examines the algorithm's resilience over time. When detecting a semi-stationary target, body movements introduce additional variations in CSI matrices, which challenge the algorithm's performance \cite{ding_2024_robust}. Therefore, evaluating the algorithm's effectiveness at different time periods becomes crucial. Experiments were conducted in the meeting room with Tp2, where the original distance between transceivers was six meters. The test setup included a row of six seats centrally located between the transceivers. Using the receiver shown in Fig.~\ref{fig:3a} as the coordinate origin, the coordinates of the six seats were (± 0.5, 3), (± 1.5, 3), and (± 2.5, 3). Reducing the number of seats to six is to minimize accumulated changes caused by chair movement. The experiments were conducted seven times (labeled as T1 to T7), with a 30-minute interval between each session, each yielding a set of ancestor sequences.

\begin{figure}[ht] 
    \centering
    \includegraphics[width=0.9\linewidth]{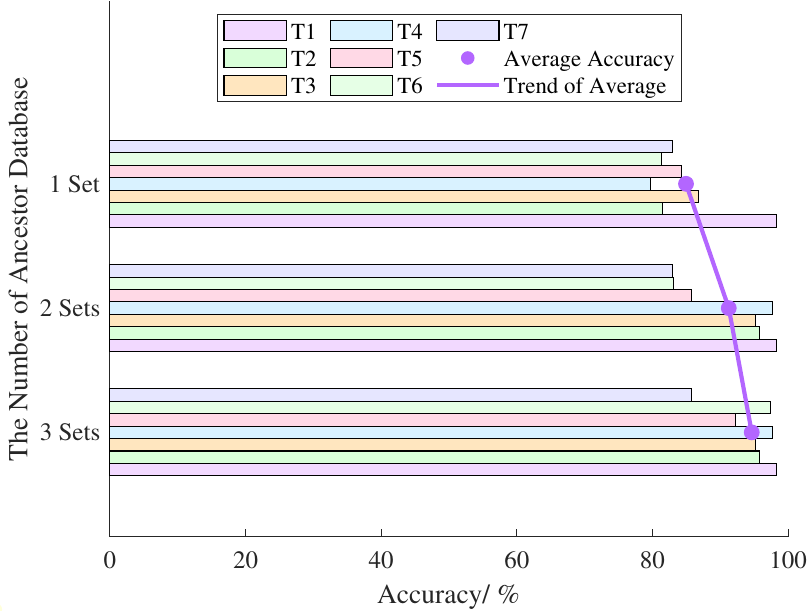}
    \caption{The accuracy of increasing with the number of sets of ancestor sequences.}
    \label{fig:5}
\end{figure}

Figure \ref{fig:5} displays the results, with the vertical axis representing the number of sets of ancestor sequences utilized. On the `One Set' vertical axis, ancestor sequences extracted from the first timestamp are used as fingerprints to match online parent sequences from subsequent timestamps. One finding is that, despite time-related fluctuations, the proposed algorithm effectively detects the position of the semi-stationary target. Although these fluctuations across different timestamps highlight the impact of time on the CSI matrix, the accuracy consistently remains at least 80\%, with an average of 85\%. Another finding is that collecting more sets of ancestor sequences further enhances the algorithm’s robustness. As depicted in Figure \ref{fig:5}, using two sets of ancestor sequences results in an average accuracy exceeding 90\%, and the average accuracy climbs to approximately 95\% when the number of ancestor sets increases to three, representing a 10\% improvement compared to using a single set. 

\subsection{Impact of Transceiver Distances}

\begin{figure}[ht] 
    \centering
    \includegraphics[width=0.9\linewidth]{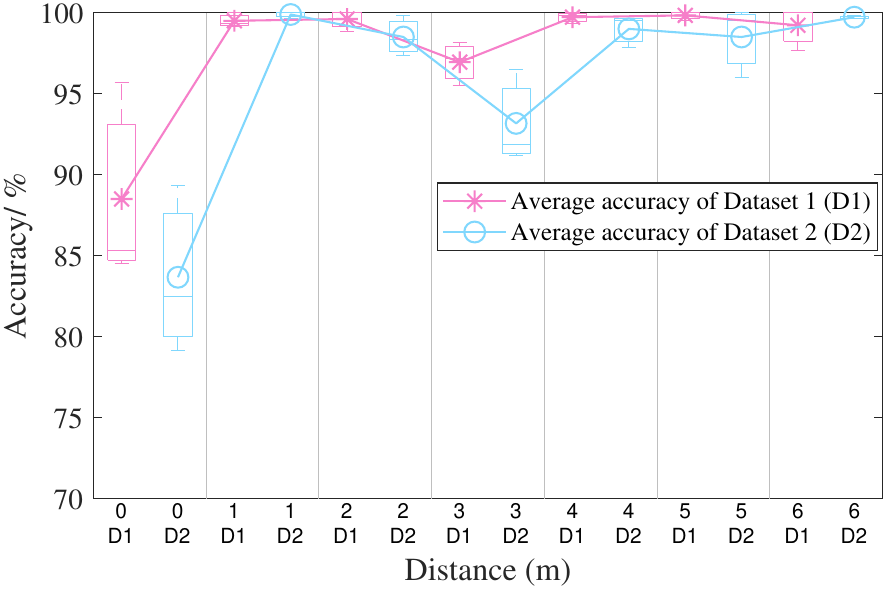}
    \caption{The accuracy of different transceiver distances.}
    \label{fig:7}
\end{figure}

This section evaluates the effects of varying distances between transceivers, utilizing the same topology outlined in Section {\uppercase\expandafter{\romannumeral 4}}.E. The experiment commenced with the initial receiver positioned at the coordinate origin, progressively moving it closer to the transmitter—from six to zero meters, reducing the distance by one meter at each step. Fig.~\ref{fig:7} displays two sets of experimental results. The average accuracy remained above 80\%, with most cases exceeding 90\%.

However, significant performance declines were observed at two specific distances. The most pronounced decline in performance occurs when the distance between transceivers is reduced to zero meters, placing them in the same spot. Based on the result in \cite{wang_2024_csibased}, the indoor Fresnel zone indicated a diminished radio sensing capability when transceivers overlap. Additionally, extending the transceiver distance to three meters leads to another drop, which is associated with interference from the target. Positioned at (0, 3), the receiver aligns with the test locations in the same row. The closer proximity makes radio waves more sensitive to the target's movements \cite{wang_2016_csibased}, leading more readily to changes in the CSI matrix. Therefore, this paper recommends maintaining a separation of at least one meter between transceivers, and between transceivers and targets, to minimize interference and optimize accuracy.

\section{CONCLUSION}

This paper introduces BiCSI, an innovative position matching algorithm designed to detect semi-stationary targets by leveraging binary encoding and fingerprinting techniques. BiCSI uniquely converts the CSI matrix into binary sequences and employs Hamming distances to measure signal similarity. In our tests, BiCSI achieves an average accuracy of above 98\% and a MAE of less than three centimeters, outperforming traditional algorithms by at least two-fold. Additionally, the results show that the binary process is capable of multiple similarity measurements, with the MAE of employing Hamming distance improving by 4\%-28\%. 

Furthermore, this work presents a novel method for extracting feature vectors from CSI matrices. Despite the latency of offline binary processes, BiCSI requires only kilobyte-scale storage, far less than the megabyte-scale storage needed by traditional machine learning models. Despite time-related fluctuations, the algorithm matches the position of the semi-stationary target with an average accuracy of 85\%, which can increase to 95\% by expanding the ancestor sequence sets from one to three. Finally, our findings suggest maintaining a separation of at least one meter between transceivers, and between transceivers and targets, to minimize interference and optimize the system performance. Overall, based on Wi-Fi and Raspberry Pi 4B, BiCSI introduces an effective approach that utilizes physical similarity measurements to detect semi-stationary targets, expanding indoor localization applications.

\section*{Acknowledgments}
This work was supported in part by the Smart Traffic Fund (Project No. PSRI/31/2202/PR) established under the Transport Department of the Hong Kong Special Administrative Region (HKSAR), China.

\bibliographystyle{IEEEtran}



\vfill

\end{document}